\documentclass[12pt]{article}
\usepackage{graphicx}
\usepackage{epsfig}
\usepackage{latexsym}
\usepackage{latexsym}
\usepackage{amssymb}
\newcommand{\labell}[1]{\label{#1}}

\newcommand{\reef}[1]{(\ref{#1})}

\def\via{{\it via}}

\def\etc{{\it etc}}

\def\eg{{\it e.g.}}

\def\ie{{\it i.e.}}

\def\Tr{{\rm Tr}}

\DeclareSymbolFont{AMSb}{U}{msb}{m}{n}
\DeclareMathSymbol{\IN}{\mathbin}{AMSb}{"4E}
\DeclareMathSymbol{\IZ}{\mathbin}{AMSb}{"5A}
\DeclareMathSymbol{\IR}{\mathbin}{AMSb}{"52}
\DeclareMathSymbol{\Q}{\mathbin}{AMSb}{"51}
\DeclareMathSymbol{\II}{\mathbin}{AMSb}{"49}
\DeclareMathSymbol{\IC}{\mathbin}{AMSb}{"43}
\DeclareMathSymbol{\IP}{\mathbin}{AMSb}{"50}
\DeclareMathSymbol{\IH}{\mathbin}{AMSb}{"48}
\DeclareMathSymbol\IA{\mathalpha}{AMSb}{"41}
\DeclareMathSymbol\IS{\mathalpha}{AMSb}{"53}

\begin{document}

\newpage
\bigskip


\bigskip
\bigskip

\begin{center}
{\Large \bf Clearing the Throat:}\\
\bigskip
{\Large \bf Irrelevant Operators and Finite Temperature}\\
\bigskip
{\Large \bf in Large $N$ Gauge Theory}

\end{center}
\bigskip
\bigskip
\bigskip

\centerline{\bf Nick Evans{$^\flat$},
 Clifford V. Johnson$^\natural$, Michela Petrini$^\sharp$}

\bigskip
\bigskip
\bigskip

\centerline{\it $^\flat$Department of Physics, University of
Southampton} 
\centerline{\it Southampton SO17 1BJ, U.K.}
\centerline{\small \tt n.evans@hep.phys.soton.ac.uk}
\centerline{$\phantom{and}$} 

\centerline{\it $^\natural$Centre for
Particle Theory, Department of Mathematical Sciences}
\centerline{\it University of Durham, Durham, DH1 3LE, U.K.}
\centerline{\small \tt c.v.johnson@durham.ac.uk}
\centerline{$\phantom{and}$} \centerline{\it$^\sharp$Centre de Physique Th{\'e}orique, Ecole Polytechnique}
\centerline{\it F-91128 Palaiseau cedex, France}


\centerline{\small \tt  michela.petrini@cpht.polytechnique.fr}

\bigskip
\bigskip


\begin{abstract}
  \vskip 2pt We study the addition of an irrelevant operator to the
  ${\cal N}=4$ supersymmetric large~$N$ $SU(N)$ gauge theory, in the
  presence of finite temperature, $T$. In the supergravity dual, the
  effect of the operator is known to correspond to a deformation of
  the AdS$_5\times S^5$ ``throat'' which restores the asymptotic ten
  dimensional Minkowski region of spacetime, completing the full
  D3--brane solution. The system at non--zero $T$ is interesting,
  since at the extremes of some of the geometrical parameters the
  geometry interpolates between a seven dimensional spherical
  Minkowskian Schwarzschild black hole (times $\IR^3$) and a five
  dimensional flat AdS Schwarzschild black hole (times $S^5$).  We
  observe that when the coupling of the 
  operator reaches a critical value, the deconfined phase, which is
  represented by the geometry with horizon, disappears for all
  temperatures, returning the system to a confined phase which is
  represented by the thermalised extremal geometry.
\end{abstract}
\newpage
\baselineskip=18pt
\setcounter{footnote}{0}


\section{Introduction}

A number of important spacetime solutions of general relativity and
string theory have the interesting feature that they have an infinite
``throat'' spacetime at their core, connected to the asymptotically
flat region by an interpolating region or ``mouth''. A simple
prototype of this is the extremal Reissner--Nordstrom black hole in
four dimensions, at whose core there is AdS$_2\times S^2$, the
Bertotti--Robinson universe\cite{bertotti}. This behaviour persists to
various spacetime solutions of string and M--theory, such as
magnetically charged black
holes\cite{Gibbons:ih,Gibbons:1987ps,Garfinkle:qj}, various
branes\cite{exactthroats2} such as three--branes\cite{blackp} and
M--branes\cite{mbranes}, and heterotic or symmetric
five--branes\cite{rey,exactthroats}.

\begin{figure}[ht]
 \begin{center}
   \scalebox{0.35}[0.35]{\includegraphics{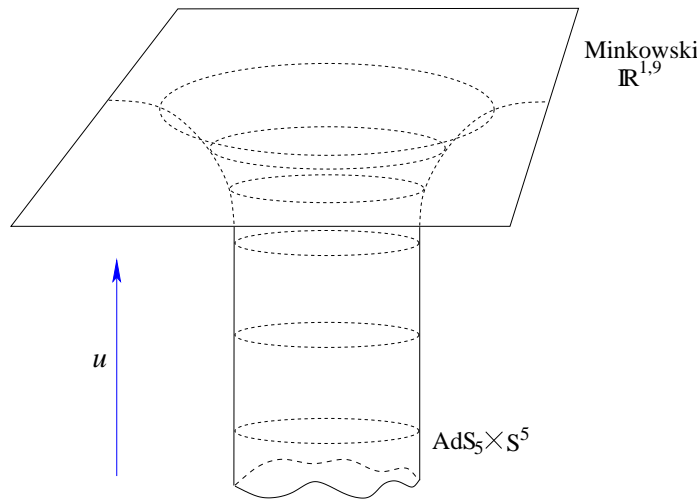}}
 \end{center}
 \caption{\small \footnotesize{The generic ``throat$+$mouth'' geometry
   of some string backgrounds of interest. We have labelled it as
   appropriate to the three--brane case, for later discussion.}}
 \label{throatpic}
 \end{figure}

The throat regions, usually possessing a high degree of symmetry, have
been of some considerable interest in string theory, since they are
often realisable as exact solutions, and some even have
representations as interesting conformal field theories (CFT's). For
example, a two dimensional black hole can be written as an exact
CFT\cite{Witten:1991yr}, using a gauged Wess--Zumino--Novikov--Witten
(WZNW) model. By tensoring this with other CFT's or embedding it
into larger gauged WZNW models to obtain non--trivial mixing between
the radial and angular sectors, this sort of model forms part of the
exact CFT description of the throat limit of various higher
dimensional objects (\eg, fivebranes\cite{exactcft}, four dimensional
magnetic holes\cite{exact4d}, four dimensional dyonic and/or rotating
holes and Taub--NUT spacetimes, \etc\cite{exact4d2}.)  Throats are
hence  important examples of how non--trivial backgrounds are
represented in string theory. See figure~\ref{throatpic}.

 A significant step in the direction of maturity of this approach was
 made when the AdS$_5\times S^5$ throat at the core of the geometry of
 $N$ coincident D3--branes\cite{blackp} was
 identified\cite{malda,gkp,w1} as being a dual description of ${\cal
   N}=4$ supersymmetric $SU(N)$ gauge theory at large $N$.  This
 particular model makes concrete many highly non--trivial ideas about
 how a theory of quantum gravity operates (such as
 holography\cite{holograms}, UV/IR relations\cite{wittsuss}), provides
 us with new tools for testing ideas about strongly coupled gauge
 theories, and may even be thought of as the best understood (at least
 in principle) non--perturbative definition of a ten dimensional
 superstring theory.
 
 One technical feature of interest in representing these throat
 geometries in string theory has been the issue of how to describe
 ---directly within the field theory representation itself--- the
 process of moving out of the throat back to the asymptotic region.
 This was once of great interest since it was hoped (for example) that
 one might be able to represent in this way the $S$--matrix for the
 scattering of quanta off a black hole, in a bid to solve the black
 hole information puzzle.
 
 In this note we study the AdS$_5\times S^5$ throat geometry,
 and its connection to the asymptotically flat region to make the
 complete D3--brane geometry. The AdS/CFT Correspondence 
identifies each supergravity field with an operator in the dual gauge theory
and so switching on these extra metric components should correspond to
some deformation of the field theory. Restoring the throat region relaxes
the decoupling limit of the AdS/CFT Correspondence and hence would
be expected to introduce gravitational and stringy couplings back into 
the field theory. We expect these effects to show up in the form of
higher dimension operators. In
 refs.\cite{perturb1,perturb2} (see also \eg, \cite{perturb3}) 
 the connection to the asymptotically flat region of the D3--brane
 geometry was identified as 
the addition of an irrelevant operator of the form
 $\Tr F^4$, which does not destroy the conformal nature of the gauge
 coupling.  Alternatively, in \cite{hash} the asymptotically flat region
was recovered in a low energy limit of a multi-centre D3 brane geometry 
describing a point on the moduli space of the ${\cal N}=4$ gauge theory.
Here the operator encodes the scalar vevs at a large scale above which the
gauge symmetry is enlarged. There are clearly many UV completions of
the theory which in the IR give rise to a higher dimension operator
of this form. Here we wish to study the role of the operator on the
system without specifying the UV completion in the spirit of, for example, the 
Nambu-Jona-Lasinio (NJL) model \cite{njl}. We will make comparisons to the 
NJL model in the final discussion. 
In particular we study the role of this higher dimension operator on the finite
 temperature behaviour of this system.  The geometry is interesting,
 since the non--extremal D3--brane geometry contains richer features
 than the extremal case.
 
 One of the most interesting of such features is that at the core
 there is the $D=5$ (flat) AdS--Schwarzschild black hole (times
 $S^5$), while asymptotically the geometry has a limit which is a
 $D=7$ asymptotically Minkowskian (spherical) Schwarzschild black hole
 (times $\IR^3$).  The relevance of the former to the dual gauge
 theory has already been demonstrated in ref.\cite{w2} as representing
 the deconfined finite temperature phase\footnote{We should clarify
   here our use of the terms ``confined'' and ``deconfined''. Of
   course, the theory really has the field content of ${\cal N}=4$
   superconformal field theory, and as such has nothing like the
   confinement that we expect from more interesting gauge theories
   with less symmetry and different field content.  The terminology
   refers to the phases observed in ref.\cite{w2} (see also
   ref.\cite{Sundborg:1999ue}): The confined phase is that which has
   zero vacuum expectation values (vevs) for the temporal Wilson line,
   while in the the deconfined phase the temporal Wilson line has a
   non--zero vev.  Such a vev measures the variation of the free
   energy due to the introduction of a static quark in the system. In
   the confined phase this is infinite and hence the vev is zero. In
   the deconfined phase the cost in free energy is finite and the vev
   is non--zero.  Another definition is directly in terms of the free
   energy~$F$: The confined phase has~$F$ of order unity, and the
   deconfined has $F$ of order $N^2$.}.  The presence of the
 irrelevant operator should therefore imply a possible role for the
 asymptotically flat $D=7$ black hole in the thermodynamics of the
 deformed gauge theory.  We allow for this possibility here.
 
 In order to make sense of these spacetimes in terms of the dual gauge
 dynamics it is important to realise that a non--renormalisable
 operator implicitly implies the existence of a cut--off, $\Lambda$,
 in the UV.  Usually such operators indicate the presence of new
 physics above $\Lambda$. We represent this in the geometries (as we
 would expect from the usual UV/IR relation) by an IR cut--off on the
 size of the space, taking the usual radial parameter out to a finite
 value, $u_{\rm max}$.  When $\Lambda$ (\ie, $u_{\rm max}$) is
 sufficiently small that it lies in the AdS part of the space, the
 physics is simply that of the usual AdS/CFT correspondence. The AdS
 black holes, which have a temperature proportional to their radius,
 are energetically preferred over a thermalised version of
 AdS\cite{w2}.
 
 In the opposite extreme where $\Lambda$ is taken large, and so the
 full asymptotically flat geometry is available, we must also consider
 the geometry which includes the Minkowskian black holes whose
 temperature is inversely proportional to their radius. Since the AdS
 and the Minkowskian black holes mutate into each other it is clear
 that there is a maximum temperature black hole geometry. At
 temperatures above this maximum value the thermalised extremal
 solution with no horizon (representing the confined phase) is the
 only possible geometry. A computation of the free energies of the
 non--extremal solution (which contains both types of black hole as
 limits) and the thermalised extremal solution below this temperature
 reveals the stronger result that the black hole solutions are {\it
   never} favoured.  This shows that the non--renormalisable operator
 destroys the deconfined phase, favouring the confined phase. When
 $\Lambda$ is taken large the non--renormalisable operator's coupling
 is large and the cut--off physics dominates. A temperature below the
 $\Lambda$ can then not change the phase of the theory.
 
 As the cut--off $\Lambda$ is reduced, making the available space more
 AdS--like, the non--renormalisable operator's coupling falls (or
 since there is only one scale in the problem we may view this as
 changing the coupling at fixed $\Lambda$). At a critical value for
 the coupling the behaviour switches between the two limits with AdS
 black holes becoming the favoured high temperature phase. Below this
 value the coupling at the cut--off is sufficiently small so as to be
 irrelevant to the IR dynamics.
 
 Note that when $\Lambda$ is taken large, the Minkowski black holes,
 whose radii are very close to the cut--off have a very negative free
 energy. Naively therefore, they appear to be the dominant low
 temperature phase at high cut--off, but we discard this possibility
 for two reasons. The first is that their large size is comparable to
 $\Lambda$ (\ie, $u_{\rm max}$), suggesting that in the field theory
 such states can not be considered without including corrections from
 trans--cut--off physics.  The other is that they have negative
 specific heat and hence cannot represent a stable (in the canonical
 ensemble we consider here) dual field theory vacuum.  We expect that
 further work will find a role for them in the non--equilibrium
 dynamics of the perturbed theory, but this is not addressed in this
 paper.
 
 The structure of this short note is as follows: In the next section,
 we present a reminder of the properties of the non--extremal
 D3--brane solution, and a brief orientation on matters concerning the
 throat limits and their relation to the dual field theory.
 Section~\ref{irrel} introduces the irrelevant operator, the return to
 flat space, and reports on the Euclidean action computation which we
 perform.  We uncover the phase structure, and draw a phase diagram in
 figure~\ref{phase}, summarising the role of how the operator and the
 cut--off work together in this beyond--AdS/CFT exploration. We end
 with a brief recapitulation and discussion in the final section.

\section{The Geometry of D3--branes}

The  fields\cite{blackp}:
\begin{eqnarray}
ds^2 &=& Z_3^{-{1\over2}}(r)\left(-K(r)dt^2+dx_1^2+dx_2^2+dx_3^2\right)
+Z_3^{1\over2}(r)\left(K(r)^{-1}dr^2+r^2d\Omega_5^2\right),\ \nonumber\\
C_{(4)}&=&\alpha_3^{-1}Z_3^{-1}dt\wedge dx_1\wedge dx_2 \wedge dx_3,\ 
\labell{thebrane}
\end{eqnarray}
with
\begin{eqnarray}
Z_3&=&1+\alpha_3\frac{r_3^4}{r^4}\ ,\qquad
K=1-\frac{r_H^4}{r^4}\ ,\nonumber\\
r_3^4&=&4\pi g_sN\alpha^{\prime 2}\ ,\qquad
\alpha_3=\sqrt{1+\left(\frac{r_H^4}{2r_3^4}\right)^2}
-\frac{r_H^4}{2r_3^4} \nonumber
\end{eqnarray}
solve the following truncation of the ten dimensional type~IIB
supergravity equations of motion:
\begin{equation}
R^\mu_{\phantom{\mu}\nu}+{1\over 240}T^\mu_{\phantom{\mu}\nu}=0\ ,
\end{equation}
where
\begin{equation}
T^\mu_{\phantom{\mu}\nu}=
5G_{(5)}^{\mu\alpha\beta\gamma\delta}G_{(5)\nu\alpha\beta\gamma\delta}-
\frac{1}{2}\delta^\mu_{\phantom{\mu}\nu}G_{(5)}^2\ ,\qquad
\mbox{and}\quad 
G_{(5)}=dC_{(4)},
\end{equation}
which can be derived from the action:
\begin{equation}
I=\frac{1}{2\kappa^2}\int d^{10}\!x \sqrt{-G}
\left(R-\frac{1}{240}G_{(5)}^2 \right)\ .
\end{equation}
Newton's constant is set by $2\kappa^2\equiv 16\pi G_N=(2\pi)^7
\alpha^{\prime 4}g_s^2$ where $g_s$ is the dimensionless closed string
coupling and $\alpha^\prime$, with dimensions of a squared length,
sets the inverse string tension.
The solution has $N$ units of D3--brane charge,
$\mu_3=(2\pi)^{-3}(\alpha^{\prime})^{-2}$.

\subsection{The Extremal Limit and the Throat}
In the limit $r_H\to0$, we find that $\alpha_3\to1$, and we recover
the extremal solution preserving 16 supercharges, representing $N$
coincident BPS D3--branes lying in the $x_1,x_2,x_3$ directions. Their
low energy dynamics, $\alpha\to 0$, are captured\cite{primer} by the
$SU(N)$ gauge theory manifest in their open string description which
is useful when $g_sN<1$. The gauge coupling is given by $g^2_{\rm
  YM}=2\pi g_s$. As is well known, the limit $g_sN>1$  has a good
description in terms of the closed string fields. Writing $r_3= 
\sqrt{\alpha'} \ell$ and
$r=\alpha^\prime u$, defining a characteristic energy scale~$u$ in the
gauge theory, the closed string fields describe the smooth
supergravity geometry at the ``throat'':
\begin{equation}
\frac{ds^2}{\alpha^\prime}= \frac{u^2}{\ell^2}
\left(-dt^2+dx_1^2+dx_2^2+dx_3^2\right)+{\ell^2}\frac{du^2}{u^2}+\ell^2d\Omega_5^2\ ,
\labell{throat}
\end{equation}
which is AdS$_5\times S^5$ with characteristic length scale
$L=\ell\sqrt{\alpha^\prime}$. The radial parameter $u$ transforms with
unit mass dimension in the dual field theory\cite{malda,gkp,w1}, as
appropriate for an energy scale.  The scaled harmonic function which
appears implicitly in the above geometry $H={\ell^4}/{u^4}$,
controlling the standard D3--brane form $(\alpha^\prime)^{-1}ds^2\sim
H^{-{1\over2}}dx_\parallel^2+H^{{1\over2}}dx_{\perp}$ can be extended
to encode the insertion of ${\cal N}=4$ preserving scalar operators in
the dual gauge theory\cite{coulomb}. For example, the form:
\begin{equation}
H=\frac{\ell^4}{u^4}\left(1+ \frac{c_{ij}
 Y^{ij}}{u^2}
+\cdots \right)
\end{equation}
encodes the insertion of the dimension $\Delta=2$ operator ${\cal
  O}_{ij}\sim \Tr[\Phi_{(i}\Phi_{j)}] $, with vacuum expectation value
$c_{ij}$, which is in the ${\bf 20'}$ of the $SO(6)\simeq SU(4)$
R--symmetry, made of the symmetric product of the six adjoint scalars
$\Phi_i$, $i=1,..,6$, in the gauge multiplet. The
$Y^{ij}(\theta,\phi,\ldots)$ schematically represent the $S^5$
spherical harmonics in the ${\bf 20'}$.  The limit of large $u$ is the
UV in the gauge theory, and we see that the operator becomes
increasingly small, as appropriate for a relevant deformation.

\subsection{Non--Extremality and Finite Temperature}
\label{hotAdS} There are two natural candidate geometries which we
might study as representing the gravity dual of the gauge theory at
finite temperature\cite{w2}. One is AdS$_5\times S^5$ itself, now
filled with quantum fields at temperature $T$, and the other is the
AdS$_5$ black hole (times $S^5$) with a flat horizon.  They are both
candidates since they are asymptotically AdS$_5\times S^5$, which is
appropriate since temperature may be represented as a relevant
operator in a low energy effective action. The black hole solution may
be obtained by taking the same throat limit as before, but of the
non--extreme solution given in equation~\reef{thebrane}, to give:
\begin{equation}
\frac{ds^2}{\alpha^\prime}=
-\left(\frac{u^2}{\ell^2}-\frac{u_H^4}{\ell^2 u^2}\right)dt^2+\frac{u^2}{\ell^2}
\left(dx_1^2+dx_2^2+dx_3^2\right)+\left(\frac{u^2}{\ell^2}-\frac{u_H^4}{\ell^2 u^2}
\right)^{-1}{du^2}+\ell^2d\Omega_5^2\ ,
\labell{flathole}
\end{equation}
where now $\ell^2=\alpha_3^{1\over2}r_3^2 / \alpha'$, and
$u_H=r_H/\alpha^\prime$. The temperature of this solution is naturally
determined by requiring regularity of the Euclidean section, with the
result\cite{euclid}
\begin{equation}T\equiv\beta^{-1}=\frac{u_H}{\pi \ell^2}\ ,
\labell{tempone}
\end{equation}
and this must be compared to AdS$_5\times S^5$ thermalised at the same
temperature (defined for example by compactifying the Euclidean
time coordinate\cite{euclid}).

\begin{figure}[ht]
 \begin{center}
 \scalebox{0.45}[0.45]{\includegraphics{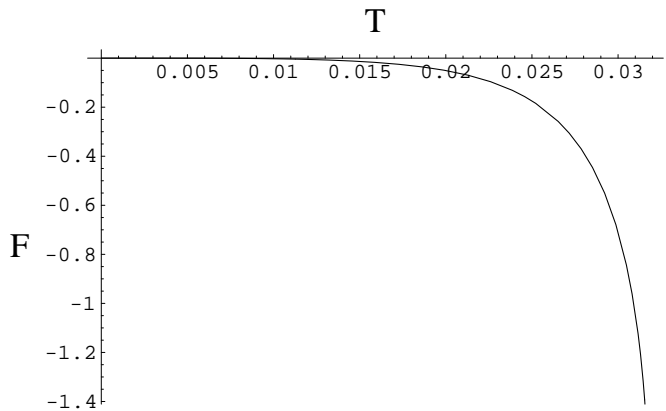}}
 \end{center}
 \caption{\small   \footnotesize 
   The thermodynamic potential $W$, or free energy $F$ of the AdS$_5$
   black holes (relative to AdS$_5$) as a function of temperature $T$.
   It is negative for all $T$, showing that these black holes dominate
   the thermodynamics, forming the deconfined phase for all $T$.}
 \label{freeAdS}
 \end{figure}

We must let the thermodynamics choose between the two solutions. We
can define for this system a canonical thermodynamic ensemble at
finite temperature $T$ by continuing the action~$I$ in the path
integral $Z$ to a Euclidean one ${\hat I}=-iI$, \via\ $t\to-i\tau$,
defining a partition function using a periodicity, $\beta$, of $\tau$
\begin{equation}
Z=\int {\cal D}[G] \exp{(-{\hat I})}=\sum_n \exp{(-\beta E_n)}\ .
\end{equation}
A careful computation in the semiclassical approximation reveals that
the free energy $F=W={\hat I}/\beta$ of AdS$_{5}$--Schwarzschild is
lower than that of AdS$_5$ for all $\beta$, (see
figure~\ref{freeAdS}), showing that it is favoured as the geometry
representing the finite temperature phase\cite{w2} which is
deconfined. This computation is the low cut--off limit of the more
general calculation we present in the next section.

\section{An Irrelevant Operator}
\label{irrel}
Consider modifying the harmonic function $H$ as follows:
\begin{equation}
H=\frac{\ell^4}{u^4}\left(\mbox{g} u^4 + 1
+\cdots \right)\ ,
\labell{modify}
\end{equation}
and we will ignore all other operators in the discussion that follows.
The coupling ${\rm g}$ (with mass dimension of ${-4}$) couples to a
new operator, ${\cal O}$, which therefore has a mass dimension of
${8}$, and has no R--charge. As discussed in the introduction a number
of UV completions of this higher dimension operator can be envisaged.
In \cite{perturb1,perturb2} the operator was identified with
$\Tr F^4$ or $(\Tr F^2)^2$, or a mixture of the two. In \cite{hash} 
it was shown that the geometry could be embedded as a limit of
a multi-centre D3 geometry and in that case the operator represents
scalar vevs at some high scale. 
If the geometry is taken in isolation, without embedding in a multi-centre
solution, then supergravity is {\it not} decoupled completely from the
theory, almost by definition: we are connecting the throat, and hence
the branes, back to the asymptotic regime where they can communicate
by exchanging gravitons, \etc, with the outside world. Gravitational effects
may therefore also contribute to the higher dimension operator.
So we have in
mind that this single operator represents the leading effect of having
integrated out the physics of the full theory in the UV but will not
specify that theory explicitly in the analysis to follow.

The existence of the
non--renormalisable operator in the field theory invites us to include
a UV cut--off, $\Lambda$, at which new physics responsible for the
operator is present. A UV cut--off in the field theory corresponds to
an IR cut--off, $u_{\rm max}$, on the radius of the geometry. We may
also think of changing the cut--off as equivalent to changing the
coupling ${\rm g}$ at a fixed $\Lambda$ since there is only one new
scale in the theory. We can see that as we raise $\Lambda$ the
operator's effects grow, with the space including more of the
asymptotic Minkowski geometry, consistent with the operator becoming
more strongly coupled.

Geometrically, the deformation corresponds to
adding the part of the harmonic function which connects AdS$_5\times
S^5$ to asymptotically flat spacetime, allowing us to move ``clear''
of the throat region. In the field theory a mass scale has been
introduced corresponding to (roughly) the radial distance at which the
two geometries interchange.  This is natural from the point of view of
the full solution given in equation~\reef{thebrane}, and the coupling
can be seen to scale as ${\rm g} \sim(\alpha^{\prime})^2$. This term
usually vanishes in the decoupling limit of the AdS/CFT
correspondence, and adding this operator is equivalent to restoring
it.  This is highly appropriate, since adding a higher dimension
operator requires the introduction of a new dimensionful scale in the
theory, which represents the physics which has been integrated out in
forming this effective operator.  Morally, we know what physics has
been integrated out: it is the full type~IIB string theory in the
presence of three--branes.

Now that we have relaxed the $\alpha^\prime\to0$ limit, we ought to
worry about stringy corrections in $\alpha^\prime$ completely
invalidating our discussion of the supergravity solution. However, we
will still keep $\alpha^\prime$ small, and also restrict ourselves to
the limit of strong 't Hooft coupling $g_{\rm YM}^2N$ so as to keep
all higher derivative corrections (coming from curvature) under
control.

The supergravity solution is obtained by simply taking the solution of
equation~\reef{thebrane}, in the extremal limit $r_H=0$, $\alpha_3=1$,
and writing it in terms of $u$ (recall that
$\ell^2=r_3^2/\alpha^\prime$):
\begin{equation}
\frac{ds^2}{\alpha^\prime}=\left(\alpha^{\prime2}+\frac{\ell^4}{u^4}\right)^{-{1\over2}}\left(-dt^2+dx_1^2+dx_2^2+dx_3^2\right) +
\left(\alpha^{\prime2}+\frac{\ell^4}{u^4}\right)^{{1\over2}}\left(du^2+u^2d\Omega_5^2\right)
\labell{irrelevant}
\end{equation}
In the limit $\alpha^\prime\to0$, the operator vanishes and we return
to the throat solution of equation~\reef{throat}.

\subsection{Going to Finite Temperature}
\label{finite}
\begin{figure}[ht]
 \begin{center}
 \scalebox{0.5}[0.5]{\includegraphics{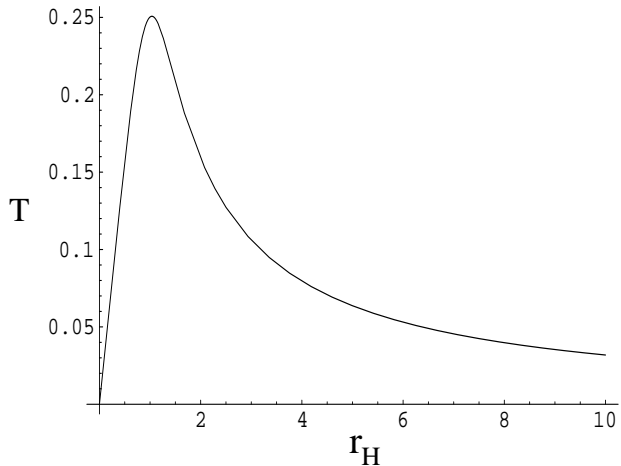}}
 \end{center}
 \caption{\small   \footnotesize The temperature of the non--extremal D3--brane solution
   as a function of $r_H$. Note two features: (1) There are two
   available solutions for each temperature, the smaller have positive
   specific heat and the larger have negative specific heat. (2) There
  is a maximum temperature beyond which there are no solutions.}
 \label{temperature}
 \end{figure}

We are interested in learning more about the role of the operator in
the theory by studying it at finite temperature. Now as before, there
are at least two candidate geometries for what might represent the
appropriate dual geometry at temperature $T$. One is the
asymptotically flat geometry, but thermalised, while the other is a
geometry with an horizon, which we write in the gauge theory
variables, starting from the brane solution in
equations~\reef{thebrane} again:
\begin{eqnarray}
\frac{ds^2}{\alpha^\prime} & = &
\bar{Z}_3(u)^{-{1\over2}}\left(- \bar{K}(u)
dt^2+dx_1^2+dx_2^2+dx_3^2\right) +
\bar{Z}_3(u)^{{1\over2}}\left(\bar{K}(u)^{-1}du^2+u^2d\Omega_5^2\right)\\ 
&& \nonumber \\
{C_{(4)} \over \alpha^{\prime 2}} & = & \alpha_3^{-1} \bar{Z}_3(u)^{-1} 
dt \wedge dx^1\wedge dx^2 \wedge dx^3
\labell{irrelevantHOT}
\end{eqnarray}
where 
\begin{equation}
\bar{Z}_3(u) = \left(\alpha^{\prime2}+\frac{\ell^4}{u^4}\right),
\hspace{1cm} \bar{K}(u) = \left(1-\frac{u_H^4}{u^4}\right)\ ,
\end{equation}
and now $\ell^2=\alpha_3^{1\over2}r_3^2/\alpha^\prime$.  This geometry
naturally has a temperature given by
\begin{equation}
\beta=  \pi r_H\left[\frac{1}{2}+\sqrt{\left(\frac{{ r}_3^4}{r_H^4}\right)^2+
\frac{1}{4}
}\right]^{1\over2} =\pi 
u_H\left[\frac{\alpha^{\prime2}}{2}
+\sqrt{\left(\frac{\ell^4}{u_H^4}\right)^2+
\frac{\alpha^{\prime4}}{4}
}\right]^{1\over2}\ ,
\labell{temptwo}
\end{equation}
obtained by the usual requirement of Euclidean regularity. Notice that
it returns to the expression~\reef{tempone} for the
AdS$_5$--Schwarzschild black hole when we switch off the operator by
sending $\alpha^\prime\to0$. We have plotted it as a function of $r_H$
in the case of non--zero $\alpha^\prime$ for later use, in
figure~\ref{temperature}.  In fact there are two classes of solutions
for each temperature, a small and large branch. The smaller solutions
are the ones which become the AdS$_5$ black holes. From the slope of
the curve, they have positive specific heat, while the larger ones,
which more closely resemble $D=7$ Schwarzschild black holes, have
negative specific heat, showing that they are unstable\footnote{This
  double set of branches of solutions at a given temperature, one
  stable and the other unstable, is similar to the case of AdS$_5$ in
  global coordinates, where the radial slices are three--spheres. There are
  spherical black holes in that case that are both large and
  small\cite{hawkpage,w2} relative to the AdS scale $\ell$.  Here, the
  AdS we have is in local coordinates, with only one allowed class of
  holes. The second branch occurs because we have the asymptotic
  region, and not just AdS$_5$.}.

 At low temperatures, for the larger branch, we see that $r_H$ (or
 $u_H$) increases and in the limit of it being large, we see from
 equation~\reef{thebrane} that $\alpha_3\to0$, $\ell \to0$, and the
 geometry becomes:
\begin{eqnarray}
ds^2&=&-\left(1-\frac{r_H^4}{r^4}\right)dt^2+dx_1^2+dx_2^2+dx_3^2 +
\left(1-\frac{r_H^4}{r^4}\right)^{-1}dr^2+r^2d\Omega_5^2\ ,
\labell{roundhole}
\end{eqnarray}
which is simply the {\it round} asymptotically flat $D=7$
Schwarzschild solution multiplied by $\IR^3$. It would be intriguing
if there was a role for this object in the context of a dual {\it
  four} dimensional gauge theory (deformed by an operator), and we
shall see shortly how it fits into the story.

\subsection{Thermodynamic Choices}
To find out which geometry is appropriate at finite temperature $T$,
we should compare the relative free energies of the two geometries.
This is an amusing computation, and it is worth describing it here.
We compute the action by continuing to Euclidean space, and writing:
\begin{eqnarray}
{\hat I}=-\frac{1}{2\kappa^2}\int_{\cal M} d^{10}\!x \sqrt{-G}
\left(R-\frac{1}{480}G_{(5)}^2 \right)-\frac{1}{ \kappa^2}
\int_{\partial{\cal M}} d^9\!x\sqrt{-h} \, {\cal K}\ .
\end{eqnarray}
where ${\cal M}$ is spacetime, ${\cal K}$ is the trace of the
extrinsic curvature tensor defined on the boundary $\partial{\cal
M}$, which here will be at our cut--off radius, and $h_{\mu \nu}$
is the nine--dimensional boundary metric.

In fact, the computation is simplified considerably by the property of
both of our solutions, the vanishing of the Ricci scalar $R$. The
$G_{(5)}$ action term may be easily evaluated:
\begin{eqnarray}
\frac{1}{\kappa^2}\int_{\cal M} d^{10}\!x \sqrt{-G}
\frac{1}{480}G_{(5)}^2 & = & { \alpha^{\prime 4}
 \over 4 \kappa^2} {\rm Vol}(S^5)
V_3 {\beta \over \alpha_3^2}  \int_{u_H}^u {x^5
\over \bar{Z}_3(x)^2}
\bar{Z}_3'(x)^{2} dx  \nonumber\\ & = & - { {\rm Vol}(S_5) V_3 \beta \over 2
\kappa^2} \frac{l^8}{\alpha_3^2} \left[ { \alpha^{\prime 2} \over x^4 \alpha^{\prime 2} 
+ \ell^4}\right]_{u_H}^u
\end{eqnarray}
where ${\rm Vol}(S^5)$ is the volume of the $S^5$ and $V_3$ is the
volume of the $\IR^3$ along the brane.

Some algebra yields the following result for the trace of extrinsic
curvature of the general solution:
\begin{equation}
{\cal K}(u)=\frac{1}{2\sqrt{G_{uu}}}G^{\mu\nu}
\frac{\partial G_{\mu\nu}}{\partial u}={1 \over 2 \alpha^{\prime 1/2}}
\frac{\bar{K}^{1/2}}{\bar{Z}_3^{1/4}}
\left[\frac{\bar{K}^\prime}{\bar{K}}
+{1\over2}\frac{\bar{Z}_3^\prime}{\bar{Z}_3}+\frac{10}{u} \right]\ ,
\labell{traced}
\end{equation}
and we also have
\begin{equation}
\sqrt{-h}= \alpha^{\prime 9/2} u^5 \bar{Z}_3^{1/4}(u)\bar{K}^{1/2}(u)\varepsilon_5\ ,
\end{equation}
where $\varepsilon_5$ is the square root of the determinant of the
metric of a round $S^5$ of unit radius. The final term in ${\cal K}$
will produce a severe divergence in the large $u$ limit, but happily
the divergence cancels in the difference between the action of the
non--extremal and the extremal solution, and so we shall avoid it.

Now the rest is straightforward, except for one important subtlety:
While the result for the integral over $\tau$ for the non--extremal
solution is simply $\beta$, given in equation~\reef{temptwo}, this
should {\it not} be used as the time period in performing the integral
over the extremal solution, if we are to compare the two accurately.
Recall that the periodicity of the time coordinate of a solution sets the
inverse temperature. However, at radius $r$, there are redshift
factors which change the temperature, coming from the fact that the
geometry is curved. If we use $\beta$ for
the non--extremal geometry, we should use $\beta^*$ for the extremal
geometry, where
\begin{equation}
\beta^*=\beta\, \bar{Z}_3^{-1/4}(u)\bar{K}^{1/2}(u)\bar{Z}_3^{1/4}(u,\alpha_3=1)\ .
\end{equation}
So we can now put this all together to compute the following
expression for the action difference at radius $u$:
\begin{eqnarray}
{\hat I} &=&-\frac{V_3}{\kappa^2}{\rm Vol}(S^5)
\Biggl[\beta\left\{
\sqrt{-h}\,{\cal K}(u) + {\ell^8 \alpha^{\prime 2}\over 2 \alpha_3^2} \left(
{ 1 \over \alpha^{\prime 2} u^4 + \ell^4} - { 1 \over \alpha^{\prime 2} u_h^4 
+ \ell^4} \right) \right\} \nonumber
\\
& &\hskip2cm -\beta^* \left\{ \sqrt{-h}\,{\cal K}(u)+ 
{\ell^8\alpha^{\prime 2}
\over 2 \alpha_3^2} \left( { 1 \over \alpha^{\prime 2} u^4 +  \ell^4} 
- { 1 \over  \ell^4}
\right) \right\}_{\alpha_3=1\atop u_H=0} \Biggr]\ ,
\end{eqnarray}
where  it is to be understood that all of the
terms in the braces in the last line are to be evaluated at
$\alpha_3=1$, $u_H=0$, since they refer to the contribution of the
extremal solution.

\begin{figure}[ht]
 \begin{center}
 \scalebox{0.6}[0.6]{\includegraphics{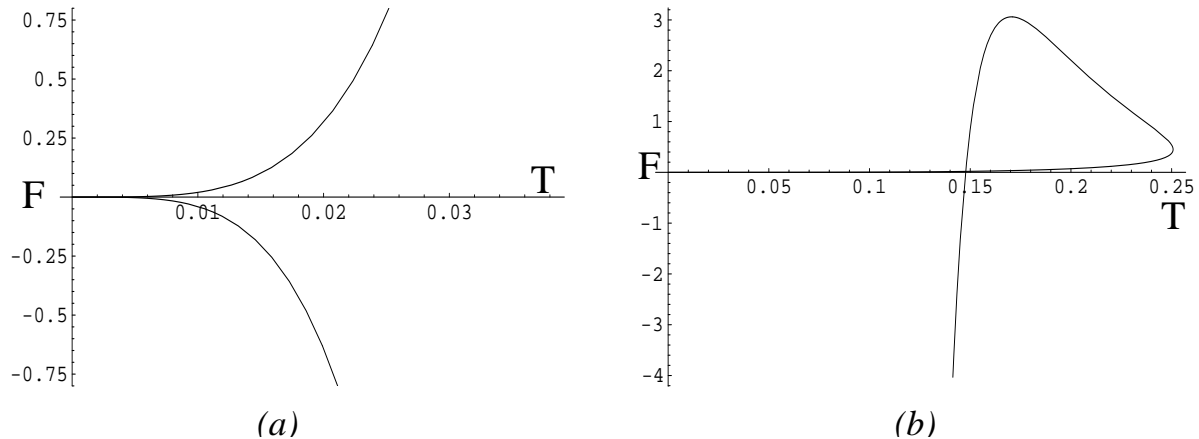}}
 \end{center}
 \caption{\small   \footnotesize
   The thermodynamic potential $W$, or free energy $F$ (relative to
   the extremal solution) as a function of temperature $T$.  {\it (a)}
   The free energy for different values of the cutoff below and above
   the critical cut--off. When the cut--off is low (bottom curve) the
   behaviour is like the pure AdS$_5$ case (see figure~\ref{freeAdS})
   with a deconfined phase while with a large cut--off (top curve) the
   confined phase is always preferred.  {\it (b)}~Large cut--off $\Lambda$:
   the snout shape is made of two branches, the steep part coming over
   the top corresponds to the large $D=7$ black holes, and the other
   branch to the smaller holes, which form the AdS$_5$ black holes in
   the throat limit. (At a certain temperature both branches merge and
   extend no further to the right.) With the large cut--off none of
   these black holes are favoured: We ignore the appearance of a
   negative free energy for the $D=7$ holes for low enough
   temperature, since they are unstable. Furthermore, at low enough
   temperatures they are simply too large for the cut--off.}
 \label{free}
\end{figure}

Since this is a rather clumsy expression, we plot it numerically in
figure~\ref{free} for varying cut--off, $u_{\rm max}$, in order to see the
result. When the cut--off is small so the space is essentially AdS the
AdS black holes are thermodynamically favoured over the thermalised
extremal solution at finite temperature. The surface term is small
(vanishing in the pure AdS limit) and the physics is dominated by the
gauge field strength term which is negative.  There is therefore a
transition from a confined phase at $T=0$ to a deconfined phase at
finite $T$.

As the cut--off is raised the positive surface term grows until at a
critical value it dominates the field strength term and the free
energy difference becomes positive. Above this critical cut--off the
thermalised extremal solution representing the confined phase is
energetically preferred. This fits our field theory intuition; as the
cut--off $\Lambda$ is raised, the non--renormalisable operator becomes
more strongly coupled at the cut--off and at some critical value we
would expect the dynamics to be controlled by that coupling at the
cut--off. Then a temperature below $\Lambda$ will not influence the
dynamics. It is interesting that the operator forces the theory back
to the confined phase.

Finally we note that when we take $\Lambda$ very large so that
Minkowskian Schwarzschild black holes are possible, their relative
free energy falls rapidly negative with decreasing $T$ and their
radius approaches the cut off scale. It would be nice to associate a
resulting role at low temperature for these $D=7$ holes. However, it
should be remembered that these holes have negative specific heat and
so are unstable in this canonical ensemble (they may simply radiate
away), and in any case their size is
sufficiently close to the cut--off that in the field theory such
states must presumably know about trans--cut--off physics which is
unspecified.  Without detailed knowledge of what lies outside the cut--off we
can not truly evaluate their role.

\begin{figure}[ht]
 \begin{center}
 \scalebox{0.5}[0.5]{\includegraphics{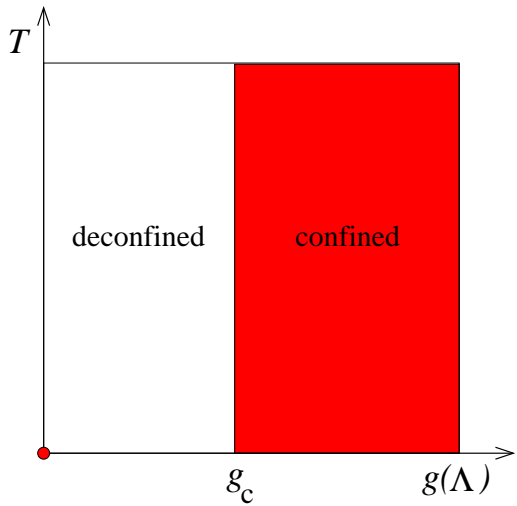}}
 \end{center}
 \caption{\footnotesize The phase diagram of the field theory as a function of
   $T$ and the coupling of the higher dimension operator.  The origin
   is in the confined phase. There is a critical value of the cutoff
   $\Lambda$ or the operator's coupling ${\rm g}$ at which there is a
   phase transition back to the confined phase. }
 \label{phase}
\end{figure}

We display the phase diagram as a function of $T$ and the
dimensionless coupling of the higher dimension operator
graphically in figure \ref{phase}.

\section{Discussion}

We have learned some interesting information about how to describe the
rest of the D3--brane supergravity geometry in the context of the
large $N$ conformal gauge theory that is dual to the theory living at
the AdS$_5\times S^5$ throat. This is quite promising, since getting
dual descriptions of the region beyond a throat (while still using the
field theory description often afforded by the throat limit)
has proven to be an arduous task in the past.

We have built onto the work of refs.\cite{perturb1,perturb2,hash}, which
suggested that this should be in terms of an irrelevant operator.
Once one leaves the throat, we can consider the full non--extremal
D3--brane geometry, and there is the potential to connect to a whole
new class of geometrical phenomena, since now we have an
asymptotically flat regime rather than asymptotically AdS. One of the
more obvious such geometries is the asymptotically flat seven
dimensional Schwarzschild black hole, with a spherical ($S^5$)
horizon, in contrast to the five dimensional AdS--Schwarzschild black
hole which has an $\IR^3$ horizon, which lives down the throat. These
two are connected, and we get a chance to see the role of the former
class in the perturbed field theory although as we have seen it is 
does not play the role of the ground state at high temperature. 
In fact that class is unstable
in the canonical ensemble\footnote{Another instability for these holes
  may be (for a range of parameters) to localise\cite{ruth, more} and
  become ten dimensional black holes. The discussion of their fate and
  role beyond that may be analogous to the discussion, presented in
  ref.\cite{gary}, of small black holes in global AdS$_5$: Possible
  stability in the micro--canonical ensemble should be considered.},
and so we do not expect it to act as a dominant controlling phase in
the thermodynamics, although it would be interesting to seek
signs of their physics in the thermodynamics of the dual 
perturbed field theory, in the light of
holography\cite{holograms}, \etc. It would also be interesting 
to study the role of these black holes with horizons close to the cut
off in the theory but this would require a fuller statement of
the UV completion of the theory. The multi-center completion \cite{hash}
where in the IR there are two copies of our solution 
appears the simplist way forward but even there complicated multi-black
hole configurations would need to be studied.

It is interesting that the irrelevant operator destroys the
deconfined phase when the cut--off is taken large. This at first
seems at odds with the intuition that an irrelevant operator
(something which seems wedded to the UV) should not encourage an
IR phenomenon such as confinement in an asymptotically free gauge
theory. However, the situation is more subtle: the operator is
evidently strongly coupled at the cut--off scale at which it is
defined, so in fact all of the
physics is determined in terms of that scale. An example to keep
in mind is the Nambu--Jona--Losino model \cite{njl}, of a free
fermion modified by a four--fermi interaction defined at some
large UV scale $M$. The resulting self--consistent solution for
the mass $m$ of the fermion is in fact of order $M$ if the
coupling is strong at the cut--off scale. If the coupling is weak
at the cut--off (or equivalently we work in the same theory but
with a lower UV cut--off) then a mass is not generated and the
operator is irrelevant to the IR physics. We have seen similar
behaviour in the case under study, now with respect to
confinement.

Finally, we close with a comment about the full string theory, which
is of course the theory that we integrated out in order to study the
conformal field theory. Our irrelevant perturbation can be thought of as
just a highly
succinct embodiment of the leading contributions of the full string
theory which was ``integrated out''. We have in mind that we must be
able to keep $\alpha^\prime$ small but not necessarily too large, and
that we can control curvatures by keeping the gauge theory strongly
coupled, so that we do not yet worry that the supergravity equations
of motion invalidate the entire set of solutions we are discussing due
to the necessity to introduce higher derivative terms. It does not
seem unreasonable to be able to relax the decoupling limit which leads
to the AdS/CFT correspondence in this way, allowing us to say more in
the context of field theory, but without having to know the details of
the entire string theory corrections.

Perhaps now that we have explored a somewhat direct route out of
the throat, and into the wider world of ten dimensional string
theory, we might be able to look afresh at many familiar geometric
solutions of string and M--theory to find useful ways of recasting
their geometrical properties which may teach us more about further
connections between gauge theory and geometry.

\section*{Acknowledgements}
We thank the Aspen Center for Physics for hospitality during the
course of this work. We thank Rob Myers for comments. N.E. is grateful
to PPARC for the sponsorship of an Advanced Fellowship. C.V.J. thanks
PPARC and The Royal Society for support and S.J.B. for her patience
and the use of a computer.  This manuscript is report \#'s SHEP-01-24,
DCTP-01/69 and CPHT-S055.1201.


\end{document}